\documentclass{nature} 

\usepackage{graphicx}
\usepackage{caption}
\usepackage{lineno}
\usepackage{authblk}
\usepackage{hyperref}
\usepackage{float}
\restylefloat{table}

\usepackage{amsmath,amssymb,amsxtra,amstext,amsfonts,dsfont}
\usepackage{color}
\usepackage[normalem]{ulem} 
\usepackage{multirow}
\usepackage{array}
\usepackage{upgreek}
\usepackage{mathrsfs}
\usepackage{mathtools}

\usepackage[squaren]{SIunits}

\newcommand{\ket}[1]{\ensuremath{|{#1}\rangle}}

\newcommand{\abs}[1]{\ensuremath{\left|{#1}\right|}}
\title{A mechanical qubit}

\author[1, 2, $\dagger$]{Yu Yang}
\author[1, 2, $\dagger$, *]{Igor Kladarić}
\author[1, 2]{Maxwell Drimmer}
\author[1, 2]{Uwe von L\"upke}
\author[1, 2]{Daan Lenterman}
\author[1, 2]{Joost Bus}
\author[1, 2]{Stefano Marti}
\author[1, 2]{Matteo Fadel}
\author[1, 2, **]{Yiwen Chu}

\affil[1]{Department of Physics, ETH Zürich, 8093 Zürich, Switzerland}
\affil[2]{Quantum Center, ETH Zürich, 8093 Zürich, Switzerland}

\affil[*]{ikladaric@phys.ethz.ch}
\affil[**]{yiwen.chu@phys.ethz.ch}

\affil[$\dagger$]{these authors contributed equally to this work}
\date{\today}

\begin{document}

\maketitle
\newline
\begin{abstract}
Strong nonlinear interactions between quantized excitations are an important resource for quantum technologies based on bosonic oscillator modes. However, most electromagnetic and mechanical nonlinearities arising from intrinsic material properties are far too weak compared to dissipation in the system to allow for nonlinear effects to be observed on the single-quantum level. To overcome this limitation, electromagnetic resonators in both the optical and microwave frequency regimes have been coupled to other strongly nonlinear quantum systems such as atoms and superconducting qubits, allowing for the demonstration of effects such as photon blockade\cite{birnbaum2005,lang2011} and coherent quantum protocols using the Kerr effect\cite{Kirchmair2013}. Here, we demonstrate the realization of the single-phonon nonlinear regime in a solid-state mechanical system. The single-phonon anharmonicity in our system exceeds the decoherence rate by a factor of 6.8, allowing us to use the lowest two energy levels of the resonator as a mechanical qubit, for which we show initialization, readout, and a complete set of direct single qubit gates. Our work adds another unique capability to a powerful quantum acoustics platform for quantum simulations\cite{von2024engineering,marti2023quantum}, sensing\cite{Goryachev_2014, Aggarwal_2021}, and information processing\cite{Pechal2019, Chamberland2020, kok2007linear}.

\end{abstract}
\maketitle 





The motional modes of mechanical systems are usually treated as harmonic oscillators in the quantum regime, with non-interacting quanta called phonons. Such a model is valid because the intrinsic mechanical nonlinearity of most materials is extremely weak and only plays a noticeable role in the classical limit of a large number of phonons. However, it has long been recognized that realizing coherent interactions at the single-quantum level in a bosonic mode can lead to new physical phenomena and important applications in quantum technologies, such as the preparation of nonclassical states and the implementation of quantum gates\cite{milburn1989, imamoglu1997}. For electromagnetic modes, the similarly weak intrinsic optical nonlinearity of materials has been overcome in engineered systems where interactions between photons are mediated through coupling to a strongly nonlinear medium. At optical frequencies, effects such as photon blockade\cite{birnbaum2005} and two-photon bound states\cite{firstenberg2013a, cantu2020repulsive} have been demonstrated using interactions with atoms. In the microwave regime, nonlinear effects of various forms in superconducting resonators can be engineered by incorporating Josephson junctions into the circuit and are used for encoding quantum information in complex bosonic states\cite{Heeres2017, Grimm2020, frattini2022squeezed}.

Compared to their electromagnetic counterparts, mechanical resonators have unique features such as long lifetimes\cite{MacCabe2020,gokhale2020epitaxial}, compactness, and the ability to directly couple to and therefore sense additional degrees of freedom such as gravitational forces\cite{Goryachev_2014, Aggarwal_2021}. As a result, there have been extensive recent advances in controlling the quantum states of mechanical objects using auxiliary systems such as superconducting circuits\cite{Satzinger2018, wollack2022, bild2023schrodinger}. However, coherent phonon-phonon interactions have thus far remained an important missing capability in quantum acoustics systems\cite{rips2013a, pistolesi2021, samanta2023}. This is largely due to the challenge of engineering a long-lived mechanical mode that retains its coherence when incorporated into a complex hybrid device and coupled to a less coherent, strongly nonlinear system.

In this work, we demonstrate a solid-state mechanical system in which the self-Kerr nonlinearity, or the phonon-phonon interaction strength, exceeds the decoherence rate. Moreover, we illustrate the promise of this device as a new building block for mechanics-based quantum information processing by operating it as a mechanical qubit. Specifically, we show that it simultaneously satisfies two basic criteria: first, although the nonlinearity is inherited through the mechanical mode's interaction with an electromagnetic system, the excitations of the system remain predominantly mechanical, or phonon-like, in character. Second, the anharmonicity $\alpha$ of the mode, defined as the energy difference between the $\ket{1}\rightarrow\ket{2}$ and the $\ket{0}\rightarrow\ket{1}$ transitions, is much greater than the decoherence rate, so that coherent quantum operations can be performed on the qubit states $\ket{0}$ and $\ket{1}$ without significantly exciting the $\ket{2}$ state. 

\begin{figure}
\centering
\includegraphics[width=9.5cm]{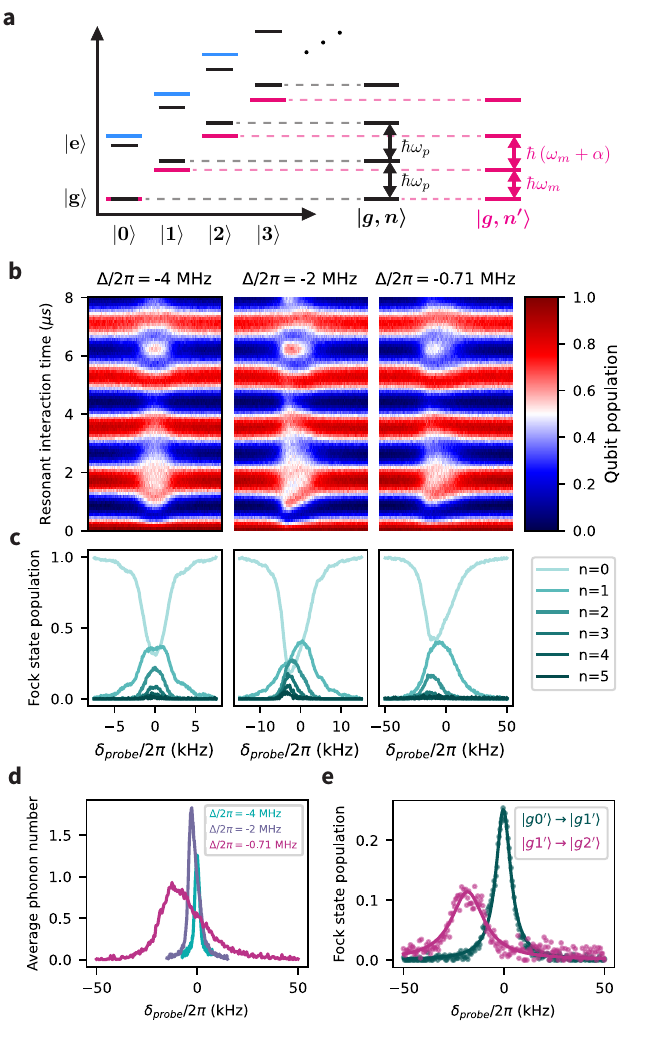}
\caption{
\linespread{1.2}\selectfont{}
\textbf{Illustration of the system energy level diagram and spectroscopic measurements.} 
\textbf{a}, Energy level diagram of the system. $\ket{n}$ denotes the $n$-phonon Fock state, while $\ket{g}$ and $\ket{e}$ represent the qubit's ground and excited states, respectively. Black horizontal lines indicate the energy levels of the bare states. Pink and blue lines represent the dressed states $\ket{gn'}$ and $\ket{en'}$, respectively. 
\textbf{b}, RPN measurement of the phonon mode as a function of probe drive frequency, for three different detunings $\Delta$. 
\textbf{c}, Phonon Fock state population as a function of probe drive frequency extracted from panel b, with which it shares the same x-axis.
\textbf{d}, Spectroscopy of the phonon mode, showing the average phonon number extracted from panel c.
\textbf{e}, Spectroscopy of the $\ket{g0'} \rightarrow \ket{g1'}$ (teal) and $\ket{g1'} \rightarrow \ket{g2'}$ (magenta) transitions, measured at $\Delta/2\pi = \unit{-0.71}{MHz}$ and fitted with Lorentzian functions.
}
\label{fig1}
\end{figure}

The device employed in our experiment is a circuit quantum acoustodynamics (cQAD) system composed of a high-overtone bulk acoustic wave resonator (HBAR) coupled to a transmon superconducting circuit, same as the one used in previous work\cite{marti2023quantum}. This allows us to use an established toolbox for quantum control and measurement to fully characterize the mechanical qubit\cite{vonLupke22}. Our system can be described by the Jaynes-Cummings (JC) Hamiltonian $H_{0}/\hbar = \omega_p p^{\dagger}p +  \frac{1}{2}\omega_q\sigma_z +g (\sigma^+ p + \sigma^- p^\dagger)$, where $\omega_q$ and $\omega_p$ are respectively the superconducting qubit and the phonon mode frequencies, $p$ is the phonon annihilation operator, $\sigma$ are the Pauli operators acting on the qubit, and $g$ is the coupling rate between the qubit and the phonon mode. Given that the anharmonicity of the superconducting qubit $\alpha_{\text{qubit}}/2\pi=-\unit{186}{MHz}$ is much larger than both the qubit-phonon detuning $\abs{\Delta}/2\pi = \abs{\omega_q - \omega_p}/2\pi \leq \unit{4}{MHz}$ and the coupling rate $g/2\pi = \unit{280}{kHz}$, we can approximate our superconducting qubit as a perfect two-level system. In the dispersive regime, where $\abs{\Delta}\gg g$, the Hamiltonian $H_0$ can be approximated by
\begin{equation}
    H_{\text{disp}}/\hbar = \omega_p p^{\dagger}p + \frac{1}{2}\left(\omega_q + \chi p^{\dagger}p\right) \sigma_z +\frac{\alpha}{2} p^{\dagger} p^{\dagger}pp\;,
    \label{eq:dispHam}
\end{equation}
where $\chi$ is the dispersive interaction between the qubit and the phonon mode\cite{vonLupke22} and $\alpha$ is the anharmonicity of the phonon mode inherited by the hybridization with the transmon qubit. The last term of Eq.~\eqref{eq:dispHam} has the form of a Kerr nonlinearity, and $\alpha$ quantifies the strength of phonon-phonon interactions in the mechanical mode. In the energy level diagram depicted in Fig.~\ref{fig1}a, $\alpha$ corresponds to the frequency difference between the $\ket{g1'} \rightarrow \ket{g2'}$ and $\ket{g0'} \rightarrow \ket{g1'}$ transitions, where $\ket{gn'}$ represents the dressed Fock state $\ket{n}$ after hybridization. It is given by (see Supplementary Information \ref{Ramsey_type_seq})
\begin{equation}
  \alpha = 
      -\frac{1}{2} \Delta \pm \frac{1}{2} \left( 2 \sqrt{\Delta^2 + 4g^2} - \sqrt{\Delta^2 + 8g^2}\right) \quad\text{when}\,\, \Delta \gtrless 0 \;,
    \label{alpha}
\end{equation}
which in both cases approximates to $2g^4/\Delta^3$ for $\abs{\Delta} \gg g$. Eq.~\eqref{alpha} shows that smaller absolute detunings lead to larger phonon anharmonicities, due to stronger hybridization between the qubit and the phonon mode. 
 
We experimentally characterize the mechanical anharmonicity by performing spectroscopic measurements of the dressed phonon states. This is achieved through probing the phonon mode at different frequencies and evaluating the resulting state by a resonant-interaction phonon number (RPN) measurement\cite{Hofheinz2008, Chu2018} (see Supplementary Information \ref{RPN_sect}). Fig.~\ref{fig1}b shows the time evolution of the qubit $\ket{e}$ population after a variable resonant interaction time with the phonon mode, following a $\unit{400}{\mu s}$-long probe drive at various detunings $\delta_{probe}$ from the dressed mechanical mode frequency $\omega_m$, corresponding to the $\ket{g0'} \rightarrow \ket{g1'}$ transition. This is done for three different qubit-phonon detunings $\Delta/2\pi=$ \unit{-4}{MHz}, \unit{-2}{MHz} and \unit{-0.71}{MHz}. From each time trace in Fig.~\ref{fig1}b, we reconstruct the phonon Fock state population distributions, which are plotted correspondingly in Fig.~\ref{fig1}c, as well as the derived average phonon number in Fig.~\ref{fig1}d.

At $\Delta/2\pi=\unit{-4}{MHz}$, where the qubit is significantly detuned from the phonon mode, the corresponding theoretical value for the anharmonicity is $\alpha_{\text{theory}}/2\pi=\unit{0.2}{kHz}$, which is considerably smaller than the phonon linewidth $\Gamma_{2}/2\pi = \unit{0.8}{kHz}$. 
When probed with a weak pulse whose driving strength is comparable with the phonon decay rate, the phonon mode behaves like an ideal harmonic oscillator. 
We refer to this as the linear resonator regime. The corresponding spectral peak has the smallest phonon linewidth in Fig.~\ref{fig1}d, and phonon populations show almost symmetric shapes as visible in Fig.~\ref{fig1}c. In the case of an intermediate detuning of $\Delta/2\pi =\unit{-2}{MHz}$, Eq.~\eqref{alpha} predicts a phonon anharmonicity $\alpha_{\text{theory}}/2\pi = -\unit{1.37}{kHz}$ that is of the same order as the phonon linewidth.
Here, the system enters the Duffing resonator regime\cite{lifshitz2008nonlinear} in which the nonlinear resonant response of the mechanical mode results in asymmetric and broadened lineshapes, as shown in Fig.~\ref{fig1}d.
The negative phonon anharmonicity causes the peaks to shift toward lower frequencies when the Fock state number increases, which can be seen in Fig.~\ref{fig1}c.
At a small detuning of $\Delta/2\pi =\unit{-0.71}{MHz}$, the calculated $\alpha_{\text{theory}}/2\pi = \unit{-17.3}{kHz}$ exceeds the total phonon linewidth $\Gamma_2/2\pi = \unit{2.52}{kHz}$. As a result, the system transitions into a qubit-like regime. The measured phonon peak width is further affected by the power broadening as shown in Fig.~\ref{fig1}d. Notably, the peak for the $n=2$ Fock state clearly appears to the left of the $n=1$ peak in Fig.~\ref{fig1}c, indicating an evident negative anharmonicity.  

In order to more precisely resolve the different transition frequencies in the qubit-like regime, we additionally probe the  $\ket{g0'} \rightarrow \ket{g1'}$ transition with a very low power drive and plot the $n=1$ population extracted by RPN measurements in Fig.~\ref{fig1}e. 
With such a weak drive, we can also resolve the $\ket{g1'} \rightarrow \ket{g2'}$ transition if there is an initial population in the $\ket{g1'}$ state. 
This is achieved by applying an additional pump drive at the $\ket{g0'} \rightarrow \ket{g1'}$ frequency while probing the $\ket{g1'} \rightarrow \ket{g2'}$ transition and extracting the resulting $n=2$ population by RPN. Using this method, we measure the transition frequency difference to be  $\unit{-17.7\pm 0.3}{kHz}$, close to the theoretically calculated $\alpha_\text{theory}/2\pi=\unit{-17.3}{kHz}$. The small discrepancy between the measurement and theory value might come from a small drift in the qubit-phonon detuning during the long measurement time.

\begin{figure}
\centering
\includegraphics[width=16cm]{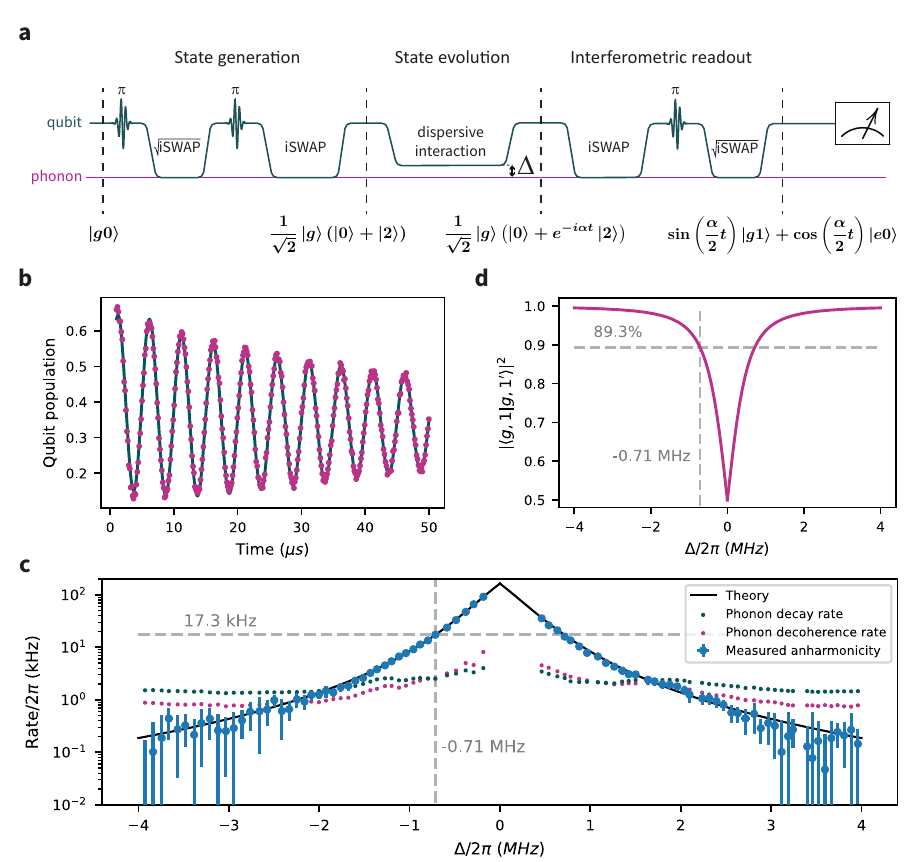}
\caption{
\linespread{1.2}\selectfont{}
\textbf{Phonon anharmonicity measurement with Ramsey-type sequence.} 
\textbf{a},  Experimental sequence for the Ramsey-type phonon anharmonicity measurement. 
\textbf{b}, Measured qubit population oscillations from the Ramsey-type anharmonicity measurement sequence. 
\textbf{c},  Measured phonon anharmonicity, decay, and decoherence rates at various qubit-phonon detunings. The theoretical value of the anharmonicity given by Eq.~\eqref{alpha} is shown in black for comparison. The measured $\alpha$ is negative for $\Delta <0$, but we plot the absolute values of $\alpha$ on a logarithmic scale for clarity. The dashed gray lines indicate the operating point for all subsequent experiments.
\textbf{d},  Calculated overlap between the dressed state $\ket{g1'}$ and the bare state $\ket{g1}$ at various qubit-phonon detunings.
}
\label{fig2}
\end{figure}

Having explored the phonon anharmonicity in the frequency domain, we perform a faster and more precise time-domain measurement of its dependence on $\Delta$ using a Ramsey-type sequence, based on the evolution of the state $\frac{1}{\sqrt{2}} \left( \ket{g0'} + \ket{g2'} \right)$. 
Fig.~\ref{fig2}a illustrates the sequence, which is comprised of three parts: state generation, state evolution, and interferometric readout. Initially, the system is in the ground state of both the qubit and phonon mode. 
A $\pi$-pulse then excites the qubit to the $\ket{e}$ state, followed by a $\sqrt{\text{iSWAP}}$ operation between the qubit and the phonon, generating the entangled state $\frac{1}{\sqrt{2}} \left( \ket{g1} + \ket{e0} \right)$. Applying another $\pi$-pulse and an iSWAP operation disentangles the qubit and the phonon mode, resulting in the superposition  $\frac{1}{\sqrt{2}} \ket{g}\left( \ket{0} + \ket{2} \right)$. We then allow the system to evolve for a time $t$ while the qubit-phonon detuning is set to $\Delta$, during which, in the rotating frame of the phonon, the state $\ket{g2'}$ will accumulate a phase $\alpha t$. 
To measure this phase, the pulse sequence used for state generation is reversed. 
The second and third $\pi$-pulses in the sequence diagram act as a reference in the Ramsey interferometer to which the phonon $\ket{g0'} \rightarrow \ket{g2'}$ transition frequency is compared. Ultimately, this results in the state $\sin(\frac{\alpha}{2} t)\ket{g1}+\cos(\frac{\alpha}{2} t)\ket{e0}$. 
To more accurately resolve the oscillation rate, we introduce an additional artificial detuning to the final $\pi$-pulse (see Supplementary Information \ref{Ramsey_type_seq}). An example of the oscillations obtained by the Ramsey-type anharmonicity measurement is shown in Fig.~\ref{fig2}b. 

We perform phonon anharmonicity measurements using the Ramsey-type sequence at various qubit-phonon detunings $\Delta$ and plot the results in Fig.~\ref{fig2}c, alongside theoretical values given by Eq.~\ref{alpha}. 
The measured anharmonicity values closely match the theoretical predictions, demonstrating the effectiveness of the Ramsey interference sequence and illustrating the tunability of the phonon anharmonicity in our system. The error bars are obtained by performing the same measurement ten times and taking the standard deviation to take into account the qubit frequency drifts between the measurements .

We additionally measure the phonon decay and decoherence rates at various detunings, as depicted in Fig.~\ref{fig2}c, for comparison with the corresponding phonon anharmonicity values. 
The measurements indicate that phonon anharmonicity, decay rate, and decoherence rate increase as the absolute detuning decreases. 
Notably, however, the anharmonicity increases more rapidly, scaling with $\Delta^{-3}$, while both phonon decay and decoherence rate are influenced by the inverse Purcell effect of the hybridization with the qubit. This inherited inverse Purcell decoherence rate $\Gamma_{2,\text{Purcell}}\approx\frac{g^2}{\Delta^2}\gamma_2$ scales as $\Delta^{-2}$ \cite{reagor2016quantum}, where $\gamma_2$ is the qubit decoherence rate. 
Consequently, the phonon anharmonicity exceeds the phonon decay and decoherence rates at sufficiently small detunings, which is a crucial condition for operating a mechanical qubit. 
Furthermore, we confirm using an analytic calculation that at such detunings, the phonon mode remains predominantly mechanical in nature, as shown in a plot of the contribution of the phonon component to the dressed state $\ket{g1'}$ in Fig.~\ref{fig2}d (see Supplementary Information \ref{hybridization}). 
At a detuning of $\Delta/2\pi=\unit{-0.71}{MHz}$, the phonon component constitutes $89.3\%$ of the dressed state, while the measured value of $\alpha/2\pi = \unit{-17.3 \pm 0.2}{kHz}$ exceeds both the phonon decay rate $\unit{2.54}{kHz}$ and decoherence rate $\unit{2.52}{kHz}$ and agrees well with $\alpha_{\text{theory}}/2\pi = \unit{-17.3}{kHz}$. All of the following experiments are conducted at this detuning, where the phonon anharmonicity is approximately 6.8 times greater than the phonon decoherence rate.

The ability to enter a regime where the phonon anharmonicity significantly exceeds the phonon decoherence rate while the system remains predominantly phonon-like is enabled by improvements in both qubit and phonon properties, as well as the qubit-phonon coupling strength, compared to our previous devices\cite{bild2023schrodinger}. 
At this qubit-phonon detuning, the phonon anharmonicity-decoherence ratio can be approximated as
\begin{equation}
    \frac{\alpha}{\Gamma_2}=\frac{\alpha}{\Gamma_{2,\text{Purcell}}+\Gamma_{2,\text{intrinsic}}}\approx\frac{2g\epsilon^3}{\epsilon^2\gamma_2+\Gamma_{2,\text{intrinsic}}}\, ,
    \label{alphagammaratio}
\end{equation}
where $\epsilon=\frac{g}{\Delta}$, $\Gamma_2$ is the total phonon decoherence rate, and $\Gamma_{2,\text{intrinsic}}$ is the intrinsic phonon decoherence rate. Based on our current system parameters, $\Gamma_{2,\text{Purcell}}$ and $\Gamma_{2,\text{intrinsic}}$ are comparable in magnitude, indicating that the current system is not solely limited by the intrinsic phonon coherence, but also by the qubit coherence. 

\begin{figure}[h!]
\centering
\includegraphics[width=16cm]{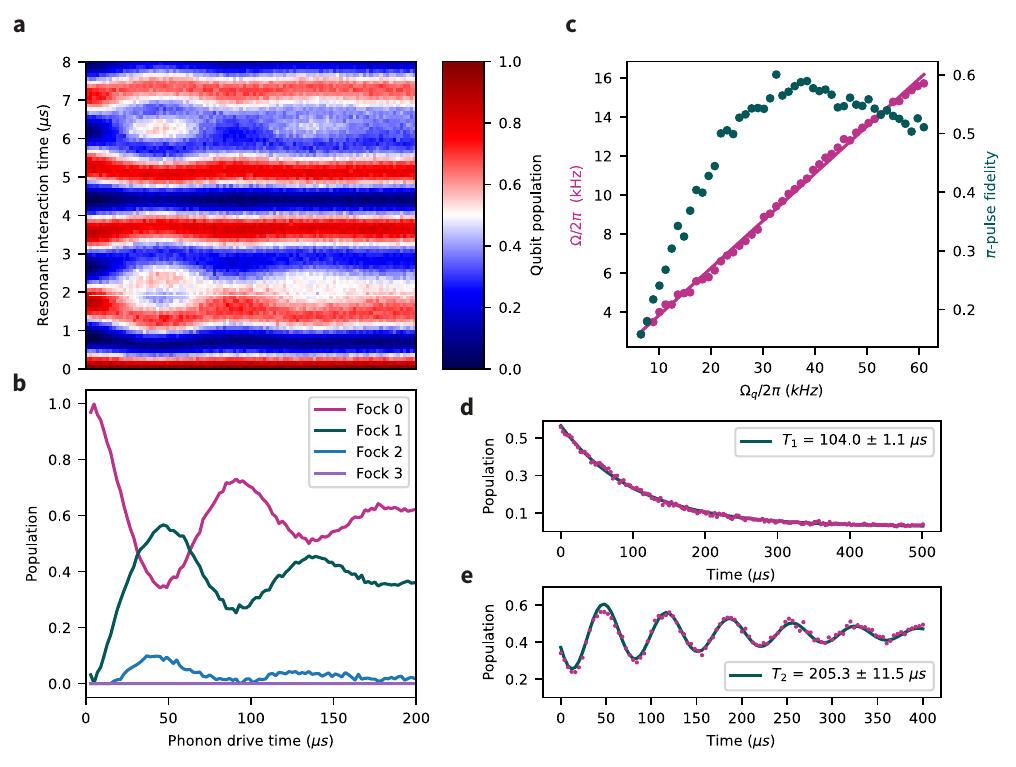}
\caption{
\linespread{1.2}\selectfont{}
\textbf{Mechanical qubit Rabi oscillations.} 
\textbf{a}, Raw RPN measurement data for $\Omega/2\pi = \unit{10.6}{kHz}$ and various drive pulse durations. 
\textbf{b}, Evolution of dressed mechanical state population distribution obtained through RPN from panel a, with which it shares the same x-axis. 
\textbf{c}, Rabi frequency (magenta) and the $\pi$-pulse fidelity (teal) as a function of the drive strength, expressed in terms of the Rabi frequency of the transmon qubit under the same drive. Dots represent measured data points, while the line is a linear fit to the measured mechanical qubit Rabi frequency.
\textbf{d, e}, Phonon $T_1$ and $T_2$-Ramsey measurement obtained using the direct mechanical qubit pulse.}
\label{fig3}
\end{figure}

We now show that our mechanical system can be operated as a qubit. Specifically, we demonstrate direct operations on the mechanical mode at a rate $\Omega$ that satisfies the condition $\Gamma_2 \ll \Omega \ll \alpha$, such that they can be seen as coherent operations on a two-level system. We first perform a direct Rabi-type measurement by applying a microwave pulse at the mechanical mode frequency of varying time $t$ and analyzing the resulting phonon Fock populations using RPN, as shown in Fig.~\ref{fig3}a. The extracted Fock state populations from each vertical slice of Fig.~\ref{fig3}a are presented in Fig.~\ref{fig3}b. 
Damped oscillations are observed between the phonon $\ket{0}$ and $\ket{1}$ state populations. Simultaneously, a small population of at most $9.4\%$ in $\ket{2}$ is also observed, indicating population leakage into higher energy states of the phonon mode. However, the population remains predominantly in the $\ket{0}$ and $\ket{1}$ states, indicating that Rabi oscillations can be performed on the mechanical system. 

We now vary the drive strength $\Omega$ and analyze the resulting Rabi oscillation frequency and the fidelity of a mechanical $\pi$-pulse, defined as the $\ket{1}$ state population at time $t_{\pi}=\pi/\Omega$ (Fig.~\ref{fig3}c). The results show that the Rabi oscillation frequency has an expected linear dependence on the drive strength. Initially, the fidelity of the $\pi$-pulse increases due to shorter operation times and smaller decoherence effects. However, the fidelity eventually reaches a maximum value and subsequently decreases, which can be attributed to the larger driving strength leading to leakage into higher Fock states. The maximum fidelity is achieved at $\Omega/2\pi=\unit{10.6}{kHz}$ for which the $\pi$-pulse fidelity is $58.9\%$ with $t_\pi=\unit{48}{\mu s}$, which is also the driving strength used in Fig.~ \ref{fig3}a, b. We continue using this $\Omega$ for all following mechanical qubit operations. For example, we perform $T_1$ and $T_2$ measurements of the mechanical mode using direct $\pi$ and $\pi/2$ mechanical qubit pulses. During the free evolution times for these measurements, we shift the qubit to a frequency far away from the mechanical mode to measure the intrinsic phonon $T_1$ and $T_2$ values (see Supplementary Information \ref{T1T2}). The results are shown in Fig.~\ref{fig3}d and \ref{fig3}e, from which we extract $T_1 = 104.0 \pm \unit{1.1}{\mu s}$ and $T_2 = 205.3 \pm \unit{11.5}{\mu s}$. These match well with values obtained by previously established techniques\cite{ChuScience2017} at large qubit-phonon detunings, as shown in Fig.~\ref{fig2}d.

\begin{figure}
\centering
\includegraphics[width=16cm]{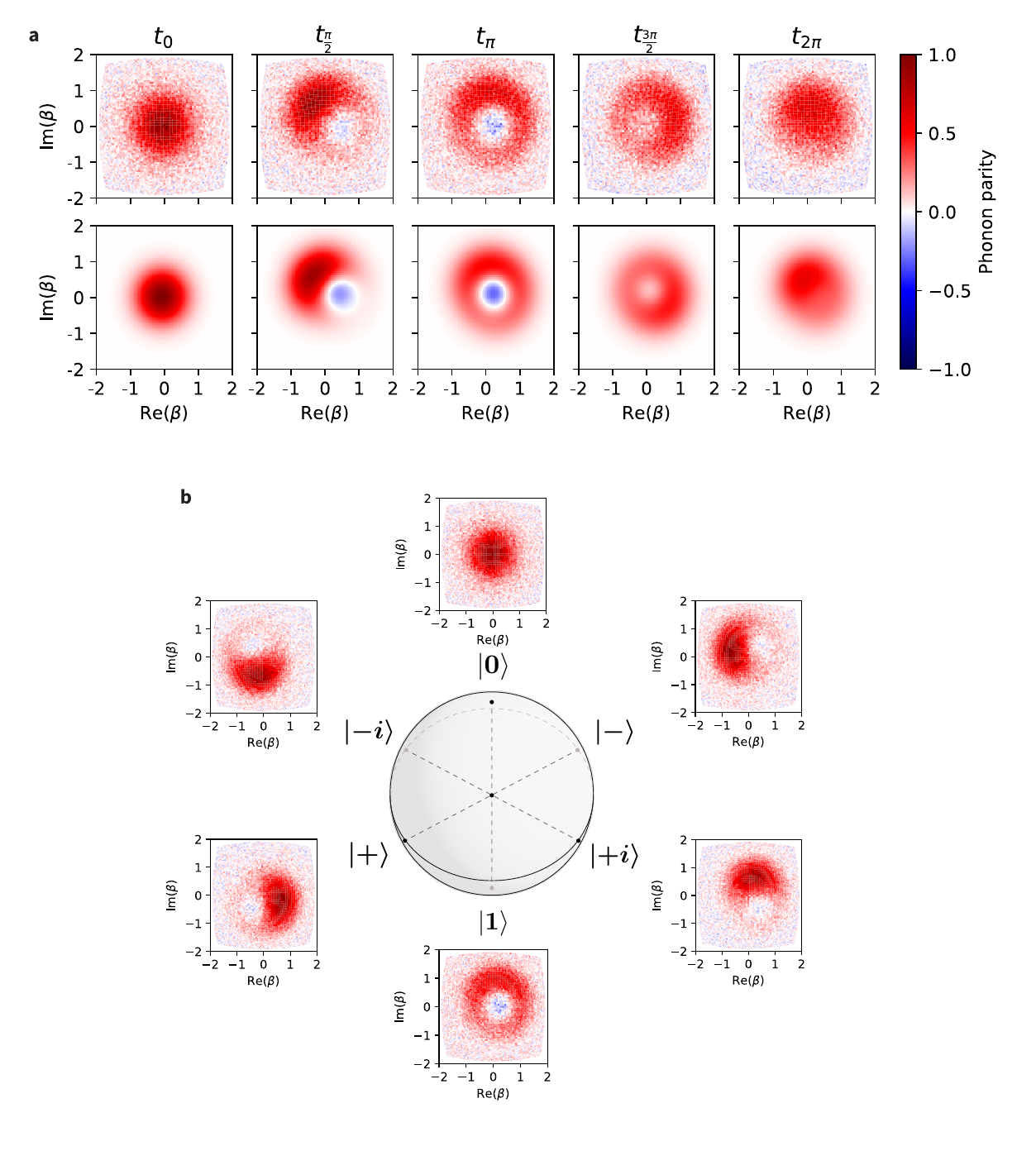}
\caption{
\linespread{1.2}\selectfont{}
\textbf{Wigner functions of the mechanical qubit states.} 
\textbf{a}, Measured (top) and simulated (bottom) Wigner functions of the phonon mode at five points during a direct mechanical qubit Rabi oscillation.
\textbf{b}, Measured Wigner functions of the phonon mode at the cardinal points of the Bloch sphere, prepared with direct mechanical qubit pulses.}
\label{fig4}
\end{figure}
One unique feature of our system is that its anharmonicity can be tuned continuously and quickly, allowing us to perform qubit operations on the mechanical system and then visualize the resulting states using Wigner tomography\cite{vonLupke22}. We demonstrate this for various driving times during a Rabi oscillation and compare the results to numerical simulations in Fig. ~\ref{fig4}a. The simulations are performed using the complete Hamiltonian of the system and take into account the system decay, decoherence, and pulse shapes, and show close agreement with the measured Wigner functions. Initially, the phonon mode is prepared in the ground state. At $t_{\frac{\pi}{2}}$, the state is approximately $\frac{1}{\sqrt{2}}(\ket{0}+e^{i\phi}\ket{1})$, where $\phi$ depends on the phase of the drive. At $t_{\pi}$, the state evolves to $\ket{1}$. Negative values in the Wigner plots also indicate that the generated state is not a coherent state resulting from driving a harmonic mode, further confirming that our mechanical system behaves as a qubit rather than a resonator. The slight offset of the Wigner function from the phase space origin can be attributed to displacement caused by partial leakage of the mechanical qubit population into higher Fock states. Further increasing the driving time to $t_{\frac{3\pi}{2}}$ does not result in a higher average phonon number, but rather the $\frac{1}{\sqrt{2}}(\ket{0}-e^{i\phi}\ket{1}$ state, which shows that the strong phonon-phonon interaction results in a phonon blockade effect. Finally, at $t_{2\pi}$, the state returns to state $\ket{0}$. The slight deviation of the final state from the ideal ground state can be explained by decoherence of the system and residual $\ket{1}$ state population. 

Finally, to complete the demonstration of full control over the mechanical qubit, we show initialization of the system at all six cardinal points on the Bloch sphere. To prepare the $\ket{+}$, $\ket{-}$, $\ket{+i}$, and $\ket{-i}$ states, the mechanical qubit is driven for a duration of $t_{\frac{\pi}{2}}$ with the phase of the driving signal set to 0$^{\circ}$, 90$^{\circ}$, 180$^{\circ}$, and 270$^{\circ}$ respectively. The resulting Wigner functions are presented in Fig.~\ref{fig4}b, which resemble the expected results, but are again of reduced contrast due to decay and decoherence of the phonon mode and slightly displaced due to leakage into higher Fock states.



Our results demonstrate a strong tunable anharmonicity of a mechanical system, achieved through dispersive coupling with a superconducting qubit. Improved device parameters allowed us to operate the mechanical system in the qubit regime, where the anharmonicity of the mechanical mode exceeds its decoherence rate. In this regime, direct operations on the mechanical qubit can be applied, whose fidelity and operation rate is limited by the anharmonicity-decoherence ratio. According to Eq.~\eqref{alphagammaratio} and in our current parameter regime, this ratio can be improved by increasing the electromechanical coupling strength or the coherence of either subsystem. Strategies for doing so include, for example, modifying the geometry of the qubit and HBAR to minimize qubit decoherence due to acoustic radiation\cite{jain2023acoustic}. In addition to improving the properties of the system, the mechanical qubit operation fidelity and operation rate can further be increased by suppressing leakage outside of the computational basis through pulse shape engineering\cite{Gambetta_2011}.

Using new physical systems to realize strong nonlinearities allows us to take advantage of their unique properties for quantum information processing. Combining our results with recently demonstrated multi-mode operations in HBAR modes\cite{von2024engineering} opens up the possibility of implementing a full set of quantum gates that do not require direct energy exchange with a superconducting qubit. This could allow us to take better advantage of long-lived mechanical modes in novel quantum computing platforms\cite{Hann2019}. Our mechanical qubit also has applications in quantum metrology, where their substantial mass makes them particularly suitable for force sensing and tests of fundamental physics\cite{Goryachev_2014, Aggarwal_2021, schrinski2023macroscopic}. Operating a mechanical system in the qubit regime also enables the use of quantum sensing protocols developed for two level systems such as atomic and spin systems\cite{du2024single,degen2017quantum}. Additionally, an adjustable strong mechanical anharmonicity expands operational possibilities not only in cQAD, but can also be utilized in optomechanical systems\cite{doeleman2023brillouin}, which potentially allows our device to be operated as a massive artificial atom with an interface to both microwave and optical light.

\section*{Acknowledgements} We thank Adrian Bachtold and Fabio Pistolesi for useful discussions. We thank Marius Bild, Simon Storz, and Alexander Grimm for comments on the manuscript. Fabrication of devices was performed at the FIRST cleanroom of ETH Zürich and the BRNC cleanroom of IBM Zürich. 

MF was supported by the Swiss National Science Foundation Ambizione Grant No. 208886. MF and SM were supported by the Branco Weiss Fellowship -- Society in Science, administered by the ETH Z\"{u}rich. YY and IK were supported by the Swiss National Science Foundation Grant No. 200021\_204073. MD was supported by the QuantERA Grant MQSens through the Swiss National Science Foundation Grant No. 20QT-1\_205542.

\section*{Author contributions}
MD, UvL, JB, YY and SM fabricated the device. YY and IK wrote experiment control sequences, performed measurements, analysed the data, and derived theoretical models. YY, IK, UvL, DL and MF performed numerical simulations of the experiments. MF and YC supervised the work. YY, IK, MF, and YC wrote the manuscript with feedback from all authors.

\section*{Competing interests} The authors declare no competing interests. \\

\section*{Data and code availability} Raw data, analysis code and QuTiP simulations are available from the corresponding authors on reasonable request.\\

\clearpage
\newpage

\renewcommand{\thesection}{}  
\renewcommand{\thesubsection}{\Alph{subsection}}  

\renewcommand{\thetable}{S\arabic{table}}  
\renewcommand{\thepage}{S\arabic{page}}  
\renewcommand{\thefigure}{S\arabic{figure}}
\renewcommand{\theequation}{S\arabic{equation}}
\setcounter{page}{1}
\setcounter{figure}{0}
\setcounter{table}{0}
\setcounter{section}{0}
\setcounter{equation}{0}

\resetlinenumber

\begin{center}
    \section*{Supplementary information for ``A mechanical qubit''}
\end{center}
    Yu Yang$^{1,2,\dagger}$, 
    Igor Kladarić$^{1,2,\dagger,*}$, 
    Maxwell Drimmer$^{1,2}$, 
    Uwe von L\"upke $^{1,2}$, 
    Daan Lenterman$^{1,2}$, 
    Joost Bus$^{1,2}$,
    Stefano Marti$^{1,2}$,
    Matteo Fadel$^{1,2}$,
    Yiwen Chu$^{1,2,**}$\\
    $^1$ \textit{Department of Physics, ETH Zürich, 8093 Zürich, Switzerland} \\
    $^2$ \textit{Quantum Center, ETH Zürich, 8093 Zürich, Switzerland} \\
    $^\dagger$ these authors contributed equally to this work\\
    $^*$ ikladaric@ethz.ch\\
    $^{**}$ yiwen.chu@phys.ethz.ch
\tableofcontents

\clearpage
\newpage
\subsection{System parameters} \label{system_params}
$\unit{}{}$
\begin{table}[h]

\centering 
\caption{\textbf{List of device parameters.} The transmon qubit properties are measured at the intrinsic frequency of the qubit without Stark shifting, and the qubit is far detuned ($\Delta\gg g$) from the phonon mode. The phonon properties are also measured when the qubit is far detuned so that the inverse Purcell effect is negligible.}
\vspace{10mm} 

\renewcommand{\arraystretch}{0.7}
\begin{tabular}{ c | c | c  }
\hline
\bf{Variable} & \bf{Parameter} & \bf{Value} \\ \hline
$\omega_q/2\pi$ & qubit frequency & $\unit{5.057}{GHz}$ \\
$T_{1,q}$ & qubit relaxation time & $\unit{23.8}{\mu s}$ \\
$T^{*}_{2, q}$ & qubit coherence time (Ramsey) & $\unit{20.4}{\mu s}$\\
$T^{E}_{2, q}$ & qubit coherence time (Echo) & $\unit{30.9}{\mu s}$\\
$\alpha_{\text{qubit}}/2\pi$ & qubit anharmonicity & $\unit{-186}{MHz}$\\
$\omega_p/2\pi$ & phonon mode frequency & $\unit{5.049}{GHz}$\\
$T_{1, p}$ & phonon mode relaxation time & $\unit{104}{\mu s}$\\
$T_{2, p}$ & phonon mode coherence time & $\unit{205}{\mu s}$\\
$g/2\pi$ & qubit-phonon coupling & $\unit{280}{kHz}$\\ \hline
\end{tabular}
\label{table:params}
\end{table}

\newpage
\subsection{Direct phonon drive} \label{phonon_drive}
\hfill\\
In our system, the phonon drive is mediated through the qubit. If we apply a drive to the qubit at the phonon frequency, the system Hamiltonian in the rotating frame of the phonon mode can be written as

\begin{equation}
    H/\hbar =\frac{1}{2}\Delta\sigma_z+g(\sigma_-p^\dagger+\sigma_+p)+\Omega_q(\sigma_-+\sigma_+)
    \;,
    \label{eq:dispHam2}
\end{equation}
where $\Omega_q$ represents the strength of the qubit drive. If we perform the Schrieffer-Wolff transformation with the unitary matrix
\begin{equation}
    U=e^{\epsilon(p^\dagger\sigma_--p\sigma_+)}\;,
\end{equation}
where $\epsilon=\frac{g}{\Delta}$, we obtain
\begin{equation}
    H'/\hbar=\frac{1}{2}\Delta\sigma_z+\Omega_q(\sigma_-+\sigma_+)
    +\epsilon\Omega_q\sigma_z(p^++p)
    +g\epsilon(p^+p-\frac{1}{2})\sigma_z+O(\epsilon^2)
    \;.
\end{equation}
The term $\epsilon\sigma_z\Omega(p^++p)$ indicates that the direct phonon drive term depends on the qubit state, and has strength $\Omega=\epsilon\Omega_q$. In this work, we only perform the direct phonon drive under the condition that the qubit is in the ground state.

\newpage
\subsection{Resonant-interaction phonon number (RPN) measurement} \label{RPN_sect}
\begin{figure}
\centering
\includegraphics[width=10cm]{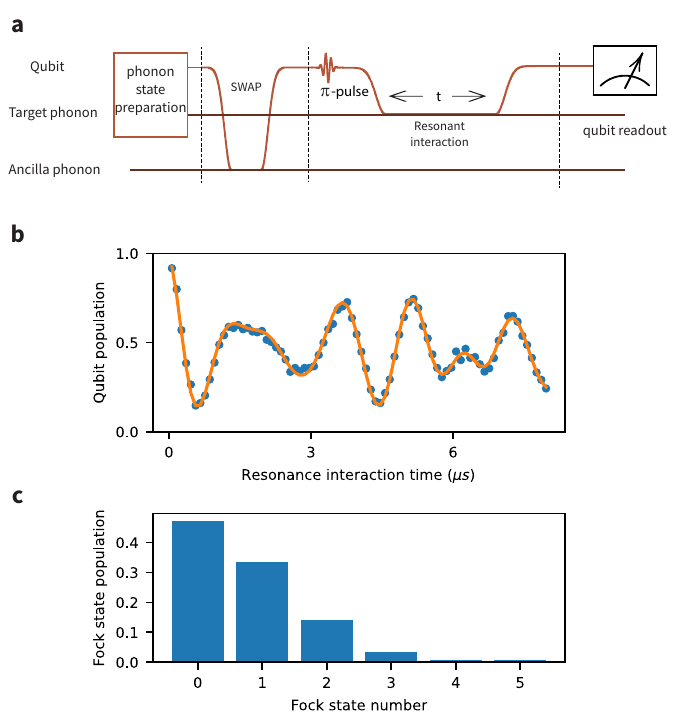}
\caption{
\linespread{1.2}\selectfont{}
\textbf{Resonant-interaction phonon number measurement.} 
\textbf{a}, Experimental sequence for the resonant-interaction phonon number measurement.
\textbf{b}, An example of measured and fitted data using RPN.
\textbf{c}, Fitted phonon population for each Fock state.
}
\label{figS1}
\end{figure}

\hfill\\
The RPN measurement has already been used in previous works\cite{Chu2018,bild2023schrodinger,von2024engineering}. We briefly describe it here for completeness. The RPN measurement sequence is illustrated in Fig.~\ref{figS1}a. First, after state preparation, we cool down and initialize the transmon qubit using a SWAP operation with an ancillary phonon mode. The qubit is then excited to the $\ket{e}$ state and subsequently tuned on resonance with the phonon mode for a time $t$. We vary $t$ and measure the qubit population, then normalize the measured data based on the readout contrast. To model this resonant interaction, we simulate the dynamics with different initial Fock states, accounting for system decay and decoherence. Using the simulated data as a basis, we identify the set of phonon populations that minimizes the residual sum of squares (RSS). This optimization includes constraints that the sum of populations in each Fock state must equal 1 and each must be within the range [0,1]. An example of measured and fitted data is displayed in Fig.~\ref{figS1}b, with the corresponding fitted Fock state populations shown in Fig.~\ref{figS1}c.

\newpage
\subsection{Ramsey-type anharmonicity measurement} \label{Ramsey_type_seq}
\subsubsection{Theory}

The Ramsey-type measurement of the anharmonicity presented in Fig.~\ref{fig2}a is based on preparing the phonon mode in the $\frac{1}{\sqrt{2}} \left( \ket{0} + \ket{2} \right)$ state, to then measure the phase $\alpha t$ accumulated during the dispersive interaction between the qubit and the phonon at various interaction times $t$. In the ideal lossless case, the resulting state is then $\frac{1}{\sqrt{2}} \left( \ket{0} + e^{-i \alpha t}\ket{2} \right)$.
 
However, for a phonon mode with decay rate $\kappa$, decoherence rate $\gamma$ and anharmonicity $\alpha$, the initial state $\frac{1}{\sqrt{2}} \left( \ket{0} + \ket{2} \right)$ evolves according to the Linblad master equation 

\begin{equation}
\frac{d\rho}{dt} = -\frac{i}{\hbar} [H, \rho] + \gamma_1 \left( p \rho p^\dagger - \frac{1}{2} \{ p^\dagger p, \rho \} \right) + 2\gamma_\phi \left( p^\dagger p \rho p^\dagger p - \frac{1}{2} \{p^\dagger p p^\dagger p, \rho\} \right) \;,
\end{equation}
where $H/\hbar = \frac{\alpha}{2} p^\dagger p^\dagger p p $, $p$ is the annihilation operator of the phonon mode, $\gamma_1$ is the phonon decay rate and $\gamma_{\phi}$ is the phonon dephasing rate. At time $t$, the initial state evolves into 
\begin{equation}
\rho \left( t \right) = \begin{pmatrix}
1+\frac{1}{2} e^{-2 \gamma_1 t} - e^{-\gamma_1 t} & 0 & \frac{1}{2} e^{-i \alpha t - \left( \gamma_1 + 4 \gamma_{\phi} \right) t}\\
0 & e^{-\gamma_1 t} - e^{-2 \gamma_1 t} & 0 \\
\frac{1}{2} e^{i \alpha t - \left( \gamma_1 + 4 \gamma_{\phi} \right) t} & 0 & \frac{1}{2} e^{-2 \gamma_1 t}
\end{pmatrix} 
\equiv \begin{pmatrix}
a & 0 & b\\
0 & c & 0 \\
d & 0 & e
\end{pmatrix}
\begin{matrix}
    \ket{0}\\
    \ket{1}\\
    \ket{2}
\end{matrix}   \;.
\end{equation}
Here, and in the following, we denote on the right side of the equation the basis of states in which the matrix elements are expressed.

During the dispersive interaction, we assume that the qubit is in the ground state. After the dispersive interaction, a series of transmon qubit and phonon operations are implemented in order to resolve the accumulated phase, see Fig.~\ref{fig2}a. After applying the iSWAP,  qubit $\pi$-pulse and $\sqrt{\text{iSWAP}}$ operations, the system ends up in the state
\begin{equation}
\rho_{\text{final}} = 
\scalebox{0.55}{$
\begin{pmatrix}
c \sin{\left( \frac{\pi}{2 \sqrt{2}} \right)^2} & 0 & \frac{c}{2} \sin{\left( \frac{\pi}{2 \sqrt{2}} \right)} \sin{\left( \frac{\pi}{\sqrt{2}} \right)} & 0 & -i \frac{c}{4} \csc{\left( \frac{\pi}{2 \sqrt{2}} \right)} \sin{\left( \frac{\pi}{\sqrt{2}} \right)^2} & 0 \\
0 & \frac{1}{2} \left( a-b-d-e \right) & 0 & -\frac{i}{2} \left( a-b-d-e \right) & 0 & 0 \\
\frac{c}{2} \sin{\left( \frac{\pi}{2 \sqrt{2}} \right)} \sin{\left( \frac{\pi}{\sqrt{2}} \right)} & 0 & \frac{c}{4} \sin{\left( \frac{\pi}{\sqrt{2}} \right)^2} & 0 & -i c \cos{\left( \frac{\pi}{2 \sqrt{2}} \right)^3} \sin{\left( \frac{\pi}{2 \sqrt{2}} \right)} & 0 \\
0 & \frac{i}{2} \left( a-b-d-e \right) & 0 & \frac{1}{2} \left( a-b-d-e \right) & 0 & 0 \\
i \frac{c}{4} \csc{\left( \frac{\pi}{2 \sqrt{2}} \right)} \sin{\left( \frac{\pi}{\sqrt{2}} \right)^2} & 0 & i c \cos{\left( \frac{\pi}{2 \sqrt{2}} \right)^3} \sin{\left( \frac{\pi}{2 \sqrt{2}} \right)} & 0 & c \cos{\left( \frac{\pi}{2 \sqrt{2}} \right)^4} & 0 \\
0 & 0 & 0 & 0 & 0 & 0
\end{pmatrix}
\begin{matrix}
    \ket{g0}\\
    \ket{g1}\\
    \ket{g2}\\
    \ket{e0}\\
    \ket{e1}\\
    \ket{e2}
\end{matrix}
$}
\end{equation}
This means that the probability of the transmon qubit being measured in the excited state is 
\begin{equation}
    \begin{aligned}
    P_e &= \frac{1}{2} \left( a-b-d+e \right) + c \cos{\left( \frac{\pi}{2 \sqrt{2}} \right)^4} \\
    &= \frac{1}{2} \left( 1 + \left( 1 - 2 \cos^4 \left( \frac{\pi}{2 \sqrt{2}} \right) \right) \left( e^{-2 \gamma_1 t} - e^{-\gamma_1 t} \right)  + e^{-\left( \gamma_1 + 4\gamma_{\phi} \right) t} \cos{\left( \alpha t \right)}\right)\,.
    \end{aligned}
    \label{full_anharmonicity_eq}
\end{equation}
By fitting this expression to the measured data, we are able to extract the anharmonicity $\alpha$.

\subsubsection{Implementation}

To accurately resolve the phonon anharmonicity with a Ramsey-type measurement sequence, we first calibrate the power of a Stark-shift drive to set the transmon qubit frequency at the targeted interaction point at which we want to measure the phonon anharmonicity. In order to obtain the real detuning between bare qubit and phonon modes, we measure the $\ket{g0'} \rightarrow \ket{g1'}$ transition frequency at this interaction point using the phonon Ramsey sequence developed in previous work\cite{Chu2018}. The real detuning $\Delta$ can then be calculated as $\Delta = \sqrt{\Delta'^2 - 4g^2}$, where $\Delta'$ is the measured frequency difference between the hybridized qubit and phonon modes.

Subsequently, the Ramsey-type anharmonicity measurement sequence is executed as presented in Fig.~\ref{fig2}a. In the experimental implementation, the final transmon qubit $\pi$-pulse is adjusted by a phase $\omega_{\text{AD}}t$ relative to the second qubit $\pi$-pulse. This additional artificial detuning $\omega_{\text{AD}} = \unit{100}{kHz}$ increases the oscillation frequency of the measured data, which enhances the final fitting precision by allowing us to better distinguish the oscillations from the decaying envelope. To determine the anharmonicity $\alpha$, we evolve the state $\frac{1}{\sqrt{2}} (\ket{g0'} + \ket{g2'})$ up to $t = \unit{50}{\mu s}$ and then fit the measurement results with the function Eq.~\eqref{full_anharmonicity_eq}, where the resulting oscillation frequency is equal to $\alpha + 2\omega_{\text{AD}}$.

\begin{figure}[h!]
\centering
\includegraphics[width=12cm]{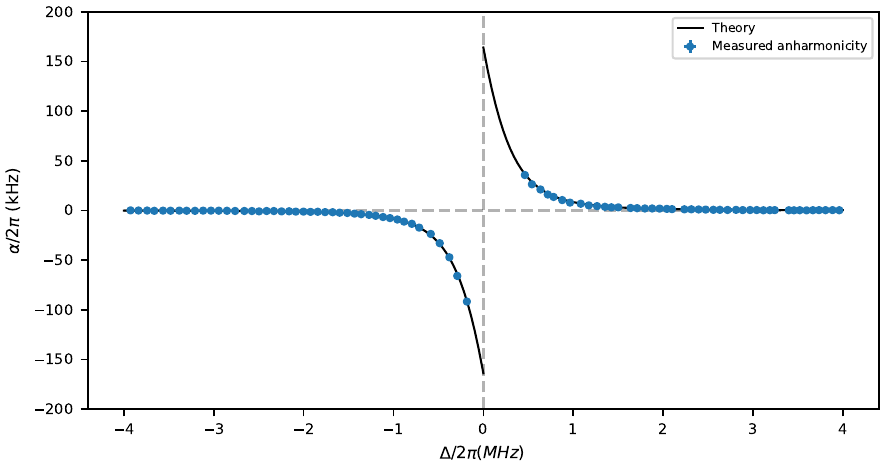}
\caption{
\linespread{1.2}\selectfont{}
\textbf{Phonon anharmonicity as a function of qubit-phonon detuning.} Data and theory are the same as Fig.~\ref{fig2}c, but shown on a linear scale.
}
\label{figS2}
\end{figure}

An example of such a measurement and its fitted function is illustrated in Fig.~\ref{fig2}b. The resulting values of the phonon anharmonicity shown in Fig.~\ref{fig2}c are shown on a linear scale in Fig.~\ref{figS2} to emphasize the ability of this Ramsey-type sequence to accurately measure both the magnitude and the sign of the induced mechanical anharmonicity.

\newpage
\subsection{Dispersive regime hybridization and anharmonicity} \label{hybridization}
\hfill\\
The Hamiltonian of the qubit-phonon system in the rotating frame of the phonon can be written as 

\begin{equation}
H/\hbar = \frac{\Delta}{2}\sigma_z + g \left( \sigma_{-} p^{\dagger} + \sigma_{+}^{\dagger} p\right) = \begin{pmatrix}
-\frac{\Delta}{2} & 0&0 &0 & 0& \\
0& \frac{\Delta}{2} & g & 0&0 & \hdots\\
0& g & -\frac{\Delta}{2} &0 &0 & \\
0& 0& 0& \frac{\Delta}{2} & g \sqrt{2} & \\
0&0 & 0& g \sqrt{2} & -\frac{\Delta}{2} & \\
& \vdots & & & & \ddots
\end{pmatrix}
\begin{matrix}
     \ket{g0}\\
     \ket{e0}\\
     \ket{g1}\\
     \ket{e1}\\
     \ket{g2}\\
     \vdots
\end{matrix} \;,
\end{equation}
where $\Delta = \omega_q - \omega_p$.
This Hamiltonian is block diagonal with blocks
\begin{equation}
H_n/\hbar = \begin{pmatrix}
\frac{\Delta}{2} & g \sqrt{n} \\
g \sqrt{n} & -\frac{\Delta}{2} \\
\end{pmatrix}
\begin{matrix}
    \ket{e, n-1}\\
    \ket{g, n}
\end{matrix}\,.
\end{equation}

For simplicity of the calculation, we will assume that $\Delta > 0$. Diagonalization of each block of the Hamiltonian allows us to obtain the expressions for the eigenstates and eigenvalues of the predominantly mechanical mode
\begin{equation}
\scalebox{0.9}{$
\ket{g, n'}\Bigm|_{\Delta>0} = 
\frac{2g \sqrt{n}}{\sqrt{4g^2 n + \left( \Delta + \sqrt{\Delta^2 + 4g^2 n} \right)^2}}  \ket{e, n-1}+
\frac{2g \sqrt{n}}{\sqrt{4g^2 n + \left( \Delta - \sqrt{\Delta^2 + 4g^2 n} \right)^2}}\ket{g, n}
$}\;,
\end{equation}
\begin{equation}
\quad E_{g,n'}\Bigm|_{\Delta>0} =  - \frac{1}{2} \sqrt{\Delta^2 + 4g^2 n}\;.
\label{Eigenenergies}
\end{equation}

The mechanical contribution to the hybridized modes can be calculated as the overlap between the dressed eigenstates and the phonon Fock states
\begin{equation}
p_{p, n}\Bigm|_{\Delta>0} = \abs{\langle g, n | g, n' \rangle }^2 = \frac{4g^2n}{4g^2 n + \left( \Delta - \sqrt{\Delta^2 + 4g^2 n} \right)^2}\;.
\label{overlap}
\end{equation}
Using the obtained expression for the energies of the hybridized mechanical modes given by Eq.~\eqref{Eigenenergies}, we can calculate the induced anharmonicity which is defined as
\begin{equation}
\alpha = \left ( E_{g,2'} - E_{g,1'}\right) - \left ( E_{g,1'} - E_{g,0'}\right)\;.
\end{equation}\ 
For positive detunings $\Delta$, the anharmonicity evaluates to
\begin{equation}
\alpha\Bigm|_{\Delta>0} = -\frac{1}{2} \Delta + \frac{1}{2} \left( 2 \sqrt{\Delta^2 + 4g^2} - \sqrt{\Delta^2 + 8g^2}\right)\;.
\end{equation}\ 

A similar approach can be used to obtain the expressions for the overlap and the anharmonicity in the case of negative detunings $\Delta$. This gives
\begin{equation}
p_{p, n}\Bigm|_{\Delta<0} = \abs{\langle g, n | g, n' \rangle }^2 = \frac{4g^2n}{4g^2 n + \left( \Delta + \sqrt{\Delta^2 + 4g^2 n} \right)^2}\;,
\label{overlap2}
\end{equation}
\begin{equation}
\alpha\Bigm|_{\Delta<0} = -\frac{1}{2} \Delta - \frac{1}{2} \left( 2 \sqrt{\Delta^2 + 4g^2} - \sqrt{\Delta^2 + 8g^2}\right)\;.
\end{equation}\

\subsection{Non-mechanical population during mechanical qubit Rabi drive} \label{squbit_pop}
\hfill\\
In order to confirm that the participation of the mechanical mode in the hybridized state evaluates to $\sim 90\text{\%}$ for the chosen qubit-phonon detuning, we measure the superconducting qubit population after the mechanical qubit drive and compare it to the value expected due to hybridization.

The participation of the bare superconducting qubit mode in the dressed mechanical mode can be calculated from Eq.~\eqref{overlap2}. For the first excited state, the participation evaluates to $p_{q,1} = 1-p_{p,1} = 10.6\text{\%}$, while for the second excited state it is $p_{q,2} = 1-p_{p,2} = 16.4\text{\%}$. A mechanical qubit $\pi$-pulse introduced in the main text produces a Fock 1 population in the mechanical state of $n^{\pi}_{p, 1} = 58.9\text{\%}$ and a Fock 2 population of $n^{\pi}_{p, 2} = 9.4\text{\%}$. Therefore, an approximation of the expected qubit population after a direct phonon $\pi$-pulse is
\begin{equation}
n^{\pi}_{q, 1} = \frac{p_{q,1}}{p_{p,1}} n^{\pi}_{p, 1} + \frac{p_{q,2}}{p_{p,2}} n^{\pi}_{p, 2} = 8.8\text{\%}\;.
\end{equation}

\begin{figure}[h!]
\centering
\includegraphics[width=10cm]{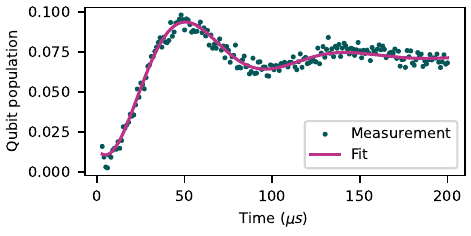}
\caption{
\linespread{1.5}\selectfont{}
\textbf{Transmon qubit population during the mechanical qubit Rabi.}
}
\label{figS3}
\end{figure}

To verify this estimate, we repeat the Rabi drive experiment presented in Fig.~\ref{fig3}a and b. However, instead of extracting the resulting phonon population using RPN, we immediately measure the superconducting qubit population. The result is shown in Fig.~\ref{figS3}. The fitting function used is of the form $A \left( 1 - e^{-\kappa t} \cos{\left( \omega t + \phi\right)} \right)$. For a $\pi$-pulse, the superconducting qubit attains a population of $n^{\pi}_{q, 1} = 9.3 \text{\%}$. We attribute the additional $0.5 \text{\%}$ in the transmon qubit population to decay of the dressed mechanical state population into the subspace of dressed superconducting qubit states during the evolution, off-resonant driving of the dressed superconducting qubit state and measurement error.

\newpage
\subsection{\boldmath Direct phonon $T_1$ and $T_2$ measurements} \label{T1T2}
\begin{figure}[h!]
\centering
\includegraphics[width=12cm]{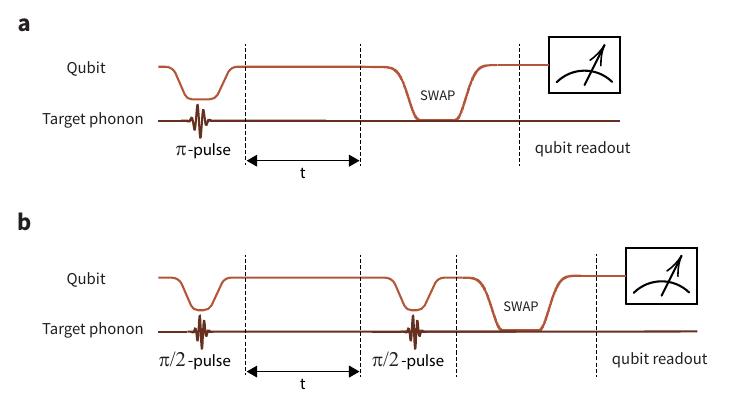}
\caption{
\linespread{1.5}\selectfont{}
\textbf{\boldmath Direct phonon $T_1$ and $T_2$-Ramsey sequences.} 
\textbf{a,} Direct phonon $T_1$ sequence.
\textbf{b,} Direct phonon $T_2$-Ramsey sequence.
}
\label{figS4}
\end{figure}

\hfill\\
To demonstrate our ability to implement direct $\pi$- and $\pi/2$-pulses on the phonon mode, we use them to conduct $T_1$ and $T_2$-Ramsey measurements on the phonon mode. Fig.~\ref{figS4}a illustrates the sequence for the $T_1$ measurement. We first shift the qubit frequency to the interaction point, which induces the phonon anharmonicity. At this point, a direct phonon $\pi$-pulse is applied to excite the phonon. Subsequently, we shift the qubit frequency away from the phonon mode, allowing the phonon to decay freely over a variable time $t$. Afterward, the remaining phonon population is measured through a swap operation with the superconducting qubit. The direct phonon $T_2$-Ramsey measurement, depicted in Fig.~\ref{figS4}b, employs a similar approach, where the $\pi/2$-pulse is applied directly to the phonon mode both before and after the free evolution time $t$.

\newpage
\subsection{State fidelity of cardinal points of the Bloch sphere} \label{Bloch_fidelities}
\hfill\\
Using a Maximum likelihood estimate\cite{chou2018}, we reconstructed the density matrix from the measured Wigner functions of the prepared states on the cardinal points of the Bloch sphere in Fig.~\ref{fig4}b. We then calculate the fidelity of the reconstructed density matrix to the target states as
\begin{equation}
    \mathcal{F}(\rho, \sigma) = \text{Tr}\left(\sqrt{\sqrt{\rho} \sigma \sqrt{\rho}}\right) \;.
\end{equation}
The resulting fidelities are $84.0\%$, $83.5\%$, $84.3\%$, $81.5\%$, $58.5\%$ for 
$\ket{-}$, $\ket{-i}$, $\ket{+}$, $\ket{i}$ and $\ket{1}$ states, respectively.

\begin{figure}[h!]
\centering
\includegraphics[width=16cm]{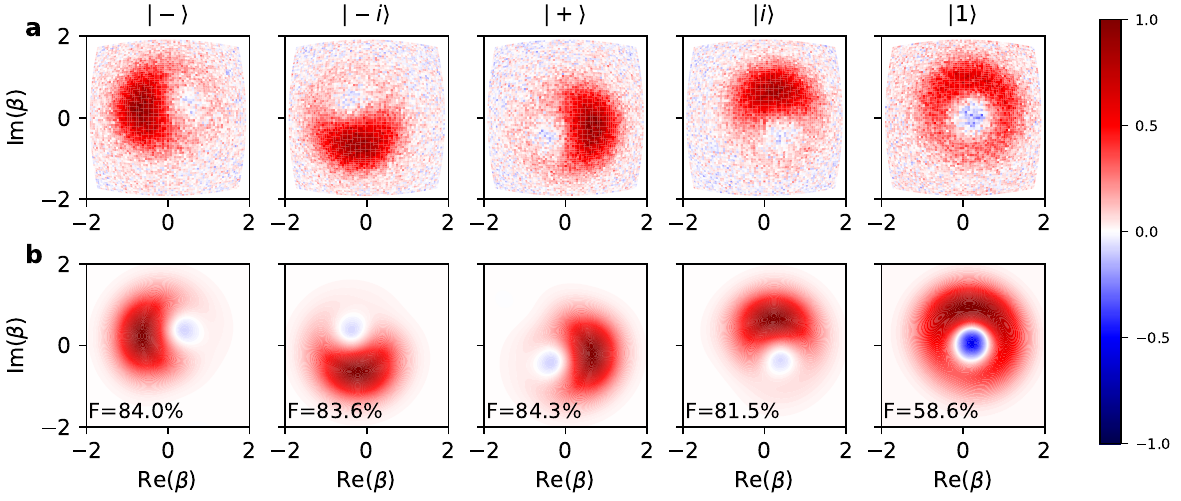}
\caption{
\linespread{1.2}\selectfont{}
\textbf{Wigner reconstruction of the prepared states.} 
\textbf{a,} Measured Wigner function of the prepared $\ket{-}$, $\ket{-i}$, $\ket{+}$, $\ket{i}$ and $\ket{1}$ states.
\textbf{b.} Wigner plot of the reconstructed density matrix. The fidelities to the target state are shown in the lower left corners.
}
\label{figS5}
\end{figure}

\newpage
\subsection{Acoustic resonator fabrication} \label{HBAR_Fab}
\hfill\\
The device used in this work was comprised of a three-dimensional transmon qubit chip bonded to an acoustic resonator chip hosting the HBAR. The qubit chip was made using identical fabrication techniques to those in Ref.\cite{Chu2018} and the bonding procedure we used is described in Ref.\cite{vonLupke22}. However, for this work, we used a modified etching process for making the acoustic resonator chip. The fabrication recipe is identical that used in our previous work\cite{Chu2018} with the following exceptions: We etch in an Oxford PlasmaPro 100 RIE/ICP etcher using a BCl$_3$/Ar plasma at 40/2 sccm, 7 mTorr, 120 W RF Power (375 V DC Bias), and 1000 W ICP power. This results in a reduced aluminum nitride etch rate of 25 nm/min and visibly cleaner domes which could be due to a lower rate of AlCl$_3$ formation\cite{pinto2022piezoelectric}, an inert byproduct of the etch chemistry. After processing, the resonator had a 0.9 nm Root Mean Square surface roughness ($S_q$) as measured by a $5\times 5$$~\mu$m atomic force microscope scan. We suspect that the low roughness and improved cleanliness may be responsible for the improvement in phonon lifetime that we observe.

\clearpage
\newpage
\bibliography{mechaqubit}

\end{document}